
\input phyzzx
\normalspace
\overfullrule=0pt
\def\delz{\partial_z}
\def\si{\sigma}
\def\am{{\alpha^\mu}}

\def\an{{\alpha}^\nu}
\def\sm #1{\sum_{n} #1}
\def\ss #1{\sum_{s} #1}
\def\e{\equiv}

\def\Om{\Omega}
\def\la{\lambda}
\def\neum #1#2{e^{{3 \atop {\sum \atop {r,s=1}}} {\infty \atop {\sum
\atop
{n,m=0}}} {1 \over 2} N_{nm}^{rs} #1 #2}}
\def\meum #1#2{e^{#1 {\infty \atop {\sum \atop {n=1}}} #2}}
\def\reum #1#2{e^{#1 {3 \atop {\sum \atop {r,s=1}}} #2}}
\def\peum #1#2{e^{#1{{3 \atop {\sum \atop {r=1}}} {\infty \atop {\sum
\atop
{n=1}}} #2}}}
\def\seum #1#2{e^{#1{{3 \atop {\sum \atop {r,s=1}}} {\infty \atop
{\sum \atop
{n=1}}} #2}}}
\def\bra #1{\left\langle#1\right\vert}
\def\ket #1{\left\vert#1\right\rangle}

\def\kin{{1 \over 2} \, (p\, +\, {Q \over 2}\, )^2}

\FRONTPAGE
\line{\hfill BROWN-HET-899}
\line{\hfill February 1993}
\bigskip
\title{{\bf
    PERTURBATION THEORY IN TWO DIMENSIONAL OPEN STRING FIELD
THEORY\foot{Work Supported in part by the Department of Energy under
contract DE-FG02-91ER40688 -- Task A.}
     }}
\bigskip
\centerline{Branko Uro\v sevi\'c}
\centerline{\it Physics Department, Brown University, Providence, RI 02912,
USA}
\bigskip
\abstract
{In this paper we develop the covariant string field theory approach
to open
$2d$ strings. Upon constructing the vertices, we apply the formalism
 to calculate the lowest order contributions to
the 4- and 5- point tachyon--tachyon tree amplitudes.
Our results are shown to
 match the 'bulk' amplitude calculations of Bershadsky and Kutasov. In
the present approach the pole structure of the amplitudes becomes
manifest and their origin as coming from the higher string modes
transparent.}

\vfill
\endpage

\chapter{Introduction}

In the last couple of years there have been flurry of activities in the
$d \le 2$ dimensional string theory. To the large extent, the
excitement was prompted by the significant success of the {\it
matrix model} approach [1-2]. Matrix model in $d=2$ have
been shown to be equivalent to the simple scalar ({\it collective})
field theory with the cubic interaction only [3].The amplitudes and
the $S$ matrix have been calculated exactly for the closed strings [4-
11].

Part of the motivation for studying the matrix models is the
hope that they can provide us with some insight about {\it string
field theory} (SFT) in higher and critical dimensions. Originally, SFT
has been formulated for the critical strings [12-18]. Covariant formulation by
Witten uses the BRST approach. It is, therefore, very
important to establish the relation between the matrix models and BRST
approaches. Partially it has been accomplished for the first
quantized BRST formulation ('Liouville theory'). The $S$-matrix
elements have been calculated, discrete states and their symmetry -
$W_{\infty}$-revealed [19-23].
Progress has been made for the open strings as well. In their paper
[24],
Bershadsky
and Kutasov calculated the bulk tree level tachyon amplitudes.
They found the pole
structure much more intriguing than in the closed string case. It is
important to note that, at present, there is no satisfactory matrix
model for the open $2d$ strings.

First quantized BRST approach corresponds to the free field theory. It
is very interesting to study the interaction theory as well. Subcritical
SFT have been discussed [25]. Some work
has been done in $d=2$ case [26-28]. One would like to: a)Solidify our
knowledge of the open SFT in $2d$; b) Check, by the explicit
calculation, that such a theory indeed reproduces the results of [24];
c) Establish the precise connection between the covariant formulation
and the simple scalar formulation in $2d$; d) After gaining some insight
dealing with the relatively simple open SFT, try to implement it to
the closed SFT, which have been recently put on the firm ground [29],
although admittedly the more complicated one. In this paper, we deal with a)
and b) leaving the rest of the program to the future investigation.

The paper is organized as follows. After the summary of our notations
and conventions (Sec.2), we discuss some general properties of the
Witten's formulation and establish the vertices  for the SFT
in $d=2$ (Sec.3). The Sec.4 is
devoted to the component analysis of the classical and quantum open
SFT. In Sec.5 we  calculate the four and five point correlation functions
using the perturbative SFT, and compare the results with [24].
Finally, we outline some open problems and possible
directions of the future work (Sec.6).

\chapter {Notations and Conventions}

In this section we summarize the notations and conventions. The matter
field $X(z)$ and the Liouville field $\varphi(z)$ are denoted as a $2d$
vector $\phi^{\mu}(z)$, where $\mu = 1,2$ corresponds
to the matter and the Liouville sector, respectively. Also:

$$\langle \phi^{\mu}(z) \phi^{\nu}(w) \rangle \sim -\delta^{\mu \nu}
\ln(z-w) \, .
\eqno\eq
$$

\noindent
The stress--energy tensor for the matter--Liouville system is given by

$$T^{\phi}(z) = -{(\delz \phi^{\mu})^2 \over 2} - i Q^{\mu} {{\delz}^2
 \phi^{\mu} \over 2}(z) ,
\eqno\eq
$$

\noindent
where $Q^{\mu}=(0, -i 2\sqrt 2)= (0, -i Q)$.

\noindent
The reparametrization ghosts are as in the critical
string case:

$$  \eqalign
{
&\langle c(z) b(w) \rangle = \langle b(z)c(w) \rangle \sim {1 \over
{z-w}}  \, , \cr
&T^{b,c}(z) = -2 b \delz c + c \delz b \, . \cr
}
\eqno\eq
$$

\noindent
In constructing the field theory vertices it is useful to bosonize ghosts.
One introduces a scalar field $\si(z)$ which 2--point function and
stress-energy tensor read:
$$  \eqalign
{
&\langle \si(z) \si(w) \rangle \sim \ln (z-w)  \, , \cr
&T^{\si} (z) = {(\delz \si)^2 \over 2} + 3{{{\delz}^2 \si} \over 2}
\, .\cr
}
\eqno\eq
$$

\noindent
Then, one can identify $c(z)\leftrightarrow e^{\si(z)}\,$,
 $b(z)\leftrightarrow e^{-\si(z)}$.
 We
will often state both the {\it bosonized} and the {\it nonbosonized}
 formulae.

\noindent
A conformal field of weight $h$, $O_h (z)$, has the
following mode expansion:

$$O_h(z) \e \sm{O_n z^{-n-h}}  \, ,
\eqno\eq
$$
with the coefficients $O_n$:
$$
O_n = \oint O(z) z^{n+h-1}\, {dz \over {2\pi i}}  \, .
\eqno\eq
$$
{}From $(2.5)$ one can read off the mode expansions for the conformal fields
$\delz \phi^{\mu} \,$, $\delz \si \,$, $c(z) \,$ and $b(z) \,$, making
use of the fact that their conformal weights are $h_{\phi} =
h_{\si}=1 \,$, $h_c=-1 \,$ and $h_b = +2$. Their expansion
coefficients are denoted, respectively, as $-i \am_n \,$, $\si_n \,$,
$c_n \,$ and $b_n$. They satisfy the (anti)commutation
relations:

$$      \eqalign{
&[\am_n \, , \an_m] = n \delta^{\mu \nu}
\delta_{n+m,0}, \cr
&[\si_n, \, \si_m] = n \delta_{n+m,0}, \cr
&\{ c_n \, , b_m \} = \{ b_n \, , c_m \} = \delta_{n+m,0} \cr    \, .
}
\eqno\eq
$$
The coefficients in the expansion of the stress--energy tensor $T(z)$
are the {\it
Virasoro} operators $L_n \,$ and they satisfy the Virasoro algebra:
$$
[L_n \, , L_m] = (n-m) L_{n+m} + {c \over 12} n(n^2-1) \delta_{n+m,0}
\, .
\eqno\eq
$$
The coefficient $c$ in $(2.8)$ is the {\it central charge} and its value
for the conformal fields we are interested in is:
$$
c_x = 1,\,\, c_{\varphi} = 1+3Q^2, \,\, c_{gh} = -26 \, .
\eqno\eq
$$

\noindent
{}From $(2.9)$ it follows that if one defines the {\it total} Virasoro
operator:
$$
L^{tot}_n \, \e L_n \e \, L^{\phi}_n + L^{gh}_n \, ,
\eqno\eq
$$
\noindent
the total central charge is $c^{tot} \e c = 0$
precisely
when $Q= 2 \sqrt2$. This specific value
of $Q$ will appear over and over again as a
consistency condition for the theory.

We define a Fock space vacuum to be the $SL(2,R)$ invariant vacuum.
This means that for  conformal field $O_h \,$ with weight $h$, the
modes $O_n$ satisfy:

$$    \eqalign
{
&O_n \ket 0 = 0  \qquad {\rm when} n \ge 1-h  \, , \cr
&\bra 0 O_{-n} = 0  \qquad {\rm when} n \ge 1-h.
}
\eqno\eq
$$

\noindent
We introduce the Hermitian conjugation as follows (here and below in
this section we closely follow the ref. [18]):

$$       \eqalign{
&(\am_n)^{\dag}\, = \am_{-n}  \qquad n \ne 0, \cr
&(\si_n)^{\dag}\, = - \, \si_{-n}  \qquad n \ne 0, \cr
&(c_n)^{\dag}\, = c_{-n}, \cr
&(b_n)^{\dag}\, = b_{-n} , \cr
&(\am_0)^{\dag}\, = \am_0 + Q^{\mu}  \qquad {\rm and} \cr
&(\si_0)^{\dag}\, = -\, \si_0 +3.  \cr
}
\eqno\eq
$$

\noindent
{}From $(2.11)$ it follows that the $SL(2,R)$ generators $\{ L_{-1}
\, , L_0 \, , L_1 \}$ annihilate both $\ket 0$ and $\bra 0$ (thus the name
for the vacuum). On the other hand, neither of the vacua is
annihilated by $c_{-1}$, $c_0$ or $c_1$ so we define:

$$
{\bra 0} c_{-1}c_0 c_1 {\ket 0}=1   \, .
\eqno\eq
$$
\noindent
This equation is the remainder of the fact that the ghost system has
the background charge $-3$ and that one therefore needs 3 zero modes to
saturate it. It is useful to introduce yet another vacuum which we refer to
as {\it physical} vacuum ${\ket \Om} $, where ${\ket \Om}  \e c_1 \ket 0$.
We postulate that $\ket 0$ has the {\it
ghost number} $g$ zero, so that ${\ket \Om}$ has $g=1$. Physical
vacuum is annihilated by {\it all} positive moded oscillators. Thus it is the
state of the lowest $L_0$ ('energy') value.
Expressed in the bosonized language: ${\ket \Om } = c_1 \ket 0 =
e^{\si(0)} \ket 0\, $. Since $[\si_0 \, , \si(z)] = +\si(z)$, one
infers that $\si_0$ is nothing but the bosonized version of the
ghost number operator $g$. More generally, one can define a state:

$$
\ket {p , \la}  =\, : e^{i p^{\mu} \phi^{\mu}(0)} e^{\la \si(0)}:
\, \ket 0,
\eqno\eq
$$
($:\, :$ denotes the normal ordering with respect to the $SL(2,R)
$ invariant vacuum). For such a state,

$$    \eqalign
{
&\am_0 \ket {p\, , \la} = p^{\mu} \ket {p\, , \la} , \cr
&\si_0 \ket {p\, , \la} = \la \ket {p\, , \la}, \cr
&\bra {p\, , \la} \am_0 = \bra {p\, , \la} p^{\mu},  \cr
&\bra {p\, , \la} \si_0 = \bra {p\, , \la} \la,  \cr
}
\eqno\eq
$$
\noindent
and the basic scalar product is defined as (note that $\bra {p\, ,
\la}$ is {\it not} the Hermitian conjugate of $\ket {p\, , \la}$):

$$
{\bra{p_1\, , \la_1}} {\ket{p_2\, , \la_2}} = {{\delta}\, (p_1 - p_2)}
{{\delta}\, (\la_1 - \la_2)} \, .
\eqno\eq
$$

\noindent
Here, $p^{\mu} \e (p, \, -iq)\, $, and $p,\, q \in R$. In the special
case when $\la = 1$, we get tachyon. To simplify the notation we use
$\ket {p\, ,1} \e \ket p\, $, unless otherwise stated. In Sec.4 and below,
'$p$' refers to the general off--shell state, while '$k$' is reserved for the
on--shell states.

\noindent
An important $g=1$ operator is the $BRST$ charge $Q$:

$$
Q = \oint :c(z) (T^{\phi} + {T^{b,c} \over 2}): \, {dz \over {2
\pi i}}.                       \eqno\eq
$$

\noindent
Making use of $(2.12)$ one easily verifies that $Q$ is
Hermitian, i.e. , that $Q^{\dag} = Q$. Also, $BRST$ charge is
nilpotent provided that the Liouville background charge is $Q = 2\sqrt 2$.
If one denotes by $F$ the 1--string Fock space built on the physical
vacuum $\ket \Om $, then $Q$ and $g$ endow $F$ with the
structure of the differential complex. Namely, $g$ provides
for the grading on the complex: $F = \oplus F^{(n)}$,
where $n$ is the ghost number of $F^{(n)}\, $, and
$Q$ is the differential. Corresponding cohomologies of $Q$, $
H^{(n)}$, are the physical states of $2d$ theory [19-23]. These two
operators play the crucial role in the
construction of the gauge invariant string field theory, as well.

\chapter{Witten's Open String Field Theory in $d=2$}

In the last section we have mentioned that the physical states of
the first quantized strings are given by the $BRST$ cohomology classes of $F$.
Transition from the first quantized to the gauge invariant field theory
formulation of the critical SFT is well known.
Namely, physical states in the first quantized approach
are the classical solutions of the free field theory
(see, eg., [17]). In [12] a covariant formulation of the interacting
SFT in $d=26$ is proposed. The present section applies the
construction to the $d=2$ case.

{\it A generic string field} $\, \ket A$ is, by
definition, an arbitrary linear combination:

$$
\ket A = \ss \ket s  a_s \, ,
\eqno\eq
$$

\noindent
where $\ket s \in F$ and $a_s$ are the coefficient functions, Grassmann
even or odd (we follow, basically, the conventions in [15] with the
shift in $g$ by $+{3 \over 2}$). The coefficient functions
depend on the center--of--mass coordinates only. Grading in the
string field space
is induced by the $F = \oplus F^{(n)}$ decomposition, so a generic
string field $\ket A$ can be decomposed into $g$--eigenstates:

$$    \eqalign
{
&\ket A = \sm {\ket A}_{(n)} \, , \cr
&g {\ket A}_{(n)} = n {\ket A}_{(n)} \e n {\ket A}_n.
}
\eqno\eq
$$

\noindent
In particular, it is convenient to declare that a Grassmann odd
coefficient functions $a_s$ {\it anticommute} with a ket
$\ket l$ if $g_l$ : $g \ket l = g_l \ket l$
is {\it odd} and that they {\it commute} otherwise. The {\it total
parity} $(-)^{g_s} \, (-)^{a_s} $ of the string field $\ket A$ is
denoted as $(-)^A$.

Next one considers multi--string states which are simply the
elements of the ${F \otimes F \cdots \otimes F} \e F^n$. A space dual to
$F^n$ we denote as $F_{ \ast n}$, where $F_{\ast n}:\,\, F^n \mapsto C$.
{\it More notations}: a mode $O_n$ corresponding
to the $r$--th string we denote as $O_n^r$; vacua are labeled as
${\ket \Om}^r$; an element from $F_{ \ast n}$ we generically denote as
$
{}_{12 \dots n}{\bra V} \e {}_n{\bra V}$. As an example,
${\ket A}_{n}^{r}$ is the $r$--th string of ghost number $n$.

Multi--string states describe the processes of splitting and joining
of strings, that is, the string interactions.
The simplest nontrivial operation is {\it integral} $\int$ which
maps $F \mapsto R$ and carries $g= -3$.
Another operation is the {\it string multiplication} {$\star$}.
Star operation carries $g=0$ and maps $F^2 \mapsto F$ so that given
the two string fields $A$ and $B$ , $A \star
B$ is again a string field. The operations $\int$ and $\star$, together with
$BRST$ charge $Q$  should satisfy Witten
axioms[12]:

$$     \eqalign
{
&(A \star B) \star C = A \star (B \star C),   \cr
&Q(A \star B) = (QA) \star B  +(-)^A A \star (QB)   \cr
&\int {A \star B} = (-)^{AB} \int {B \star A},    \cr
&\int QA = 0.
}
\eqno\eq
$$

\noindent
Using the axioms $(3.3)$ it is easy to construct a gauge invariant
theory of string fields,
classical action being:

$$
W_{cl} = {1 \over 2} \int (A_1 \star QA_1 + {2 \over 3} A_1 \star A_1
\star A_1)
\eqno\eq
$$

\noindent
and the gauge invariance is ($A_0 \e \Lambda$):

$$
\Delta A_1 = Q\Lambda + A_1 \star \Lambda - \Lambda \star A_1.
\eqno\eq
$$

\noindent
Witten's proposal can be realized by rephrasing the axioms in
terms of the Fock space oscillators [13,14]. Namely, the operations
$\int $ and the string functional multiplication $\star \, $ may be
represented through the multi--point vertex operators $\, {}_n{\bra V} \, \in
F_{\ast n}$:

$$   \eqalign
{
&\int A \e {}_1{\bra V} \ket A^1 \, , \cr
&{\ket {A \star B}}_1  \e {}_{1^{\dag}23}{\bra V}  {\ket A}^2 {\ket
B}^3 \, .
}
\eqno\eq
$$

\noindent
and the
derivative operator $Q$ is the first quantized $BRST$ operator
$(2.17)$. Then, in the bosonic string case, the classical action
$(3.4)$ reads

$$
W_{cl} = {1 \over 2} {}_{12}{\bra V} {\ket A}^2_1 Q{\ket A}^1_1 +\,  {1 \over
3} {}_{123}{\bra V} {\ket A}^3_1 {\ket A}^2_1 {\ket A}^1_1 \, ,
\eqno\eq
$$

\noindent
where $\, {}_{12}{\bra V} = {}_{123}{\bra V} {\ket V}^3 \,$ is the 2--string
vertex which is nothing but the inner product on the string
field space.

\noindent
To find ${}_{12 \dots n}{\bra V}$ is
the main problem in the construction of the theory. This can be done
by solving the {\it overlap} equations, or equivalently, using the
Neumann function method [13]. The overlap
equations for a conformal field $O_h(z)$ with conformal dimension $h$
are given by:

$$
{}_{12 \dots n}{\bra V} (z^h O^r_h(z)  - \, z^{-h} O^{r-1}_h (-z^{-1}))
= 0  \, ,
\eqno\eq
$$

\noindent
where $r=1, \dots n$, and if $r-1=0\, $ we identify
$r-1 = n$.
This leads, in particular, to the following identities for the
modes of the conformal fields in question (see Sec.2):

$$   \eqalign
{
&{}_{12 \dots n}{\bra V}\,  ({\am}^r_m + (-)^m {\am}^{r-1}_{-m}) = 0,
 \cr
&{}_{12 \dots n}{\bra V}\,  ({\si}^r_m + (-)^m {\si}^{r-1}_{-m}) = 0, \cr
&{}_{12 \dots n}{\bra V}\, (c^r_m + (-)^m c^{r-1}_{-m}) = 0, \cr
&{}_{12 \dots n}{\bra V}\, (b^r_m - (-)^m b^{r-1}_{-m}) = 0, \cr
}
\eqno\eq
$$

\noindent
where $m \ne 0$ in all four equations $(3.9)$.
Zero modes are treated separately. We require that:

$$   \eqalign
{
&{}_{12 \dots n}{\bra V}\, (\sum_{r=1}^n {\am}^r_0 + Q^{\mu}) =
{}_{12 \dots n}{\bra
V} {\delta}^{(2)}\, (\sum_{r=1}^n p^{\mu r} + Q^{\mu})=0\, , \cr
&{}_{12 \dots n}{\bra V}\, (\sum_{r=1}^n {\si}^r_0 - 3) = {}_{12 \dots n}{\bra
V} {\delta}\,(\sum_{r=1}^n {\la}^r - \, 3)\, = \, 0. \cr
}
\eqno\eq
$$

Information above is sufficient, using the results and
methods of the critical STF, to figure out, almost without further
ado, $2d$ SFT vertices. They are of the following generic form:

$$
{}_{12 \dots n}{\bra V} = {}_{12
\dots n}{\bra V^{\phi}} \,{}_{12 \dots n}{\bra V^{\si}} \, ,
\eqno\eq
$$

\noindent
where ${}_{12 \dots n}{\bra V^{\phi}}\,$ is the matter and
${}_{12 \dots n}{\bra V^{\si}}\,$ is the ghost part of
the vertex. Then, for the 1--string vertex one gets (here
$\bra p$ does not contain any ghost dependence and should not be
confused with $\bra {p,\, 1}\, $):

$$  \eqalign
{
&{}_{1}{\bra V^{\phi}} = \int \, d^2p \, {\delta}^{(2)}\, (p^{\mu} + Q^{\mu})
\bra p {\meum {-} {{(-)^n \over 2n} {\am_n} {\am_n}}}
{\meum {- Q^{\mu}} {{(-)^n \over 2n} {\am_{2n}}}},  \cr
&{}_{1}{\bra V^{\si}} = \int \, {d \la} \, {\delta} (\la - 3) {\bra
\la} {\meum {-} {{(-)^n \over 2n} {\si_n} {\si_n}}} {\meum {3} {{(-
)^n \over 2n} {\si_{2n}}}}. \cr
}
\eqno\eq
$$

For the 2--string vertex we get:

$$  \eqalign
{
&{}_{21}{\bra V^{\phi}}= \int \, {d^2}p_1 \, {d^2}p_2 \, {\delta}^{(2)}
(p^{\mu}_1 + p^{\mu}_2 + Q^{\mu}) \, {}_{2}{\bra {p_2}}\,  {}_{1}{\bra
{p_1}}
 {\meum {-} {{(-)^n \over n}
{\am_n}^1 {\am_n}^2}} \, , \cr
&{}_{21}{\bra V^{\si}}= \int \, {d {\la}_1}{d {\la}_2} \, {\delta}
({\la}_1 + {\la}_2 -3) \, {}_{2}{\bra {\la_2}} \, {}_{1}{\bra {\la_1}}
{\meum {-} {{(-)^n \over n} {\si_n}^1 {\si_n}^2}} \, .\cr
}
\eqno\eq
$$

It is useful to express the ghost part of the vertex  in the fermionic form
as well:

$$
{}_{21}{\bra V^{b,c}} = {}_{2}{\bra \Om} \, {}_{1}{\bra \Om} \, (c^1_0
+ c^2_0 ) {\meum {} {(-)^n (b^1_n c^2_n + b^2_n c^1_n)}}  \, .
\eqno\eq
$$

The 3--string vertex is:

$$   \eqalign
{
&{}_{321}{\bra V^{\phi}} = \int \, {{d^2} p_1}\, {{d^2} p_2}\, {{d^2}
p_3} \, {\delta}^{(2)} \, (p^{\mu}_1 + p^{\mu}_2 + p^{\mu}_3 +
Q^{\mu}) \, {}_{3}{\bra {p_3}} \, {}_{2}{\bra {p_2}} \, {}_{1}{\bra
{p_1}}  \cr
&{\neum {\am_n^{r}} {\am_m^{s}}}
{\peum { Q^{\mu}/3} {{(-)^n \over 2n} \am^r_{2n}}} {\reum
{Q^{\mu}/2} {N^{rs}_{00} {\am^r_0}}}  \cr
&{\seum { Q^{\mu}/3}{N^{rs}_{n0}
{\am^r_n}}} e^{-3N_{00}} \, , \cr
}
\eqno\eq
$$
$$   \eqalign
{
&{}_{321}{\bra V^{\si}} = \int \, {d \la_1}\, {d \la_2}\, {d \la_3}\,
{\delta}\, (\la_1 + \la_2 + \la_3 -3) \, {}_{3}{\bra {\la_3}}\,
{}_{2}{\bra {\la_2}}\, {}_{1}{\bra {\la_1}} \cr
&{\neum {\si^r_n} {\si^s_m}} {\peum {-} {{(-)^n \over 2n}
\si^r_{2n}}} {\reum {-} {N^{rs}_{00} {\si^r_0}}}  \cr
&{\seum {-} {N^{rs}_{n0} {\si^r_n}}} e^{3/2 {N_{00}}} \, . \cr
}
\eqno\eq
$$

\noindent
The explicit expressions for the Neumann coefficients
$N^{rs}_{nm}$ which appear in $(3.15-16)$ can be found in [18].

Note that the structure of the matter and bosonized
ghosts vertices is the same, only the values for the
insertions are different. More importantly, insertions are given,
in accordance to the geometric considerations in [12], by the
background charge of the conformal fields in question.
For the matter field $\phi^{\mu}$ it is $Q^{\mu} = (0,-i 2 \sqrt2) \,$,
and for the ghosts, it is $-3$. These are the values for which the
matter field $c$--number anomalies cancel their ghost counterparts
(this can be shown in the straightforward fashion,
just like it was done in [13] for the critical strings),
and for which the first quantized $BRST$ charge is nilpotent.

Now that we
have completed the construction of the vertices, a comment is in
order.
Liouville (non)conservation Law is explicitly enforced on the vertices
by the very construction. So, the correlation functions, calculated by
means of the Feynman rules (see Sec.5), are necessarily 'bulk'.In other
words, the cosmological constant in the theory is taken to vanish. It is
 intriguing to wander whether this condition can be relaxed and the
space--time gauge invariance preserved. An attempt in that direction
was made in [26] in which the effect of the Liouville wall was
simulated by the 'semifree' field boundary conditions. The correlation
functions calculated in that paper are, however, bulk.
Much better understanding of the cosmological constant role in the
theory is needed and will be the subject of further investigations.

\chapter {The Component Analysis in 2D SFT}

The purpose of this section is to set the ground for the explicit
 calculations of amplitudes in Sec.5. It is, also, instructive to see
how component fields enter the
classical action (Sec. 4.1) in order to better understand what happens
to them upon quantization (Sec.4.2).

Before moving on, let us introduce another piece of notation. Let us define,
for a string field $\ket A$, the reflected field
${\ket A}^r\, $ ([18,29]):
$$
{\ket A}^r_1 \, \e \, {}_{12}{\bra V} {\ket A}^2  \, .
\eqno\eq
$$
\noindent
A string field $\ket A$ is {\it real} if: ${\ket A}^r = {\ket A}^{\dag}
\,$, where $\dag \, $ denotes the Hermitian conjugation (see $(2.12)$).
{}From now on we restrict ourselves to the real string
fields (this is necessary for the
proper counting of the string degrees of freedom).

\section {Classical Field Theory in D=2}

A string field can be
represented, in general, as:
$\ket A = {\ket A}^{(N=0)} + {\ket A}^{(N=1)} + {\ket A}^{(N=2)} +
\dots$ where dots are the contributions of the levels $N \ge 3$. In this
subsection we are interested in the classical STF, so the field $\ket A$
has $g=1\, $. For such a field:

$$   \eqalign
{
&{\ket A}^{(N=0)} = \int \, {d^2 p}\,\,  \Phi (p) {\ket p}  \, ,    \cr
&{\ket A}^{(N=1)} = \int \, {d^2 p} \,\, (\, {A_{\mu} (p)} \am_{-1} {\ket p} \,
{-
i\Psi (p)} \, b_{-1} c_0 {\ket p})  \, ,   \cr
&{\ket A}^{(N=2)} = \int \, {d^2 p} \,\, (\,  -{1 \over 2} {H_{\mu \nu}(p)}
\am_{-1}
\an_{-1} {\ket p} - \, {i\, {G_{\mu} (p)} \am_{-2}\, {\ket p}} + \,  \cr
&+ \, \, {S(p)\, b_{-1} c_{-1}\, {\ket p}}\, + \, {B(p) b_{-2} c_0 {\ket p}} \,
 {-i B_{\mu} (p)\,  \am_{-1} b_{-1}
c_0\, {\ket p}} ) \, .   \cr
}
\eqno\eq
$$

\noindent
It is straightforward, although tedious, to calculate the free
field action ${1 \over 2} \, \int A\, \star
Q A\, $, which can be rewritten as:

$$
{1 \over 2} \, \int  A \, \star QA \, = \, {1 \over 2} \,
{\bra A} Q {\ket A}   \, .
\eqno\eq
$$

\noindent
On the $N=0$ level there is only one field, the tachyon:

$$
W_{f}^{(N=0)} \, = \, {1 \over 2} \, \int \, {d^2} p_1  \, {d^2} p_2 \,
{\delta}^{(2)}\, (p_1 + p_2 + Q)
\phi(p_1) (\, {p_2 \over 2}  \cdot \, (p_2 + Q)\,  - 1\, ) \, \phi
(p_2) \, .
\eqno\eq
$$

\noindent
Since ${p \over 2} \cdot (p+Q) -1 \, = \, {1 \over 2} \,
(p\, +\, {Q \over 2}\, )^2 $, one sees that tachyon in $2d$ is indeed
{\it massless}. Note that if one defines

$$
{\int}_l \, \e \, \int \,
{d^2}  p_1
d^2 p_2 \, {\delta}^{(2)}\, (p_1 +p_2 + Q) \, ,
\eqno\eq
$$

\noindent
inside the ${\int}_l\, , $  $p_1^{\mu}\,  =\,  -\, (p_2 + Q)^{\mu}$ so
$\kin$ is invariant under the partial integration.

\noindent
On the next level, $N=1$, there are two fields $A_{\mu}$ and $\Psi\, $,
the second one being an auxiliary, nondynamical field:
$$   \eqalign
{
&W_{f}^{(N=1)} \, = \, {1 \over 2} {\int}_l \, A_{\mu}(p_1)\,
(\kin +1)  A_{\mu} (p_2)\,
-i \, {\int}_l \, A_{\mu} (p_1) \, p_2^{\mu} \, \Psi (p_2) \, + \cr
&+  \,
{\int}_l \Psi\, (p_1) \Psi (p_2) \, .
}
\eqno\eq
$$

\noindent
Finally, let us state the $N=2$ action:

$$  \eqalign
{
&W_{f}^{(N=2)} \, = \, {1 \over 4}\, {\int}_l \, H_{\mu \nu} (p_1) \,
(\kin \, +2) \, H_{\mu \nu} (p_2) \,
 \cr
&- {1 \over 2}\, {\int}_l \, S(p_1) \, (\kin +2) \, S(p_2)
+\, {\int}_l \, G_{\mu}\, (p_1)\, ( \kin \, +2)\, G_{\mu}(p_2)\, +
 \cr
&+\, 2\, {\int}_l G_{\mu} (p_1) \, B_{\mu} \, (p_2) \,
+ {\int}_l  \,
B_{\mu} (p_1) B_{\mu}(p_2)\, + \, i {\int}_l \, B_{\mu}(p_1) \, p_2^{\mu}
S(p_2)\, + \,  \cr
&+ \, i {\int}_l \, B_{\mu}(p_2) \, p_2^{\nu}
H_{\mu \nu}(p_1)\,
+ \, 2{\int}_l B(p_1) \, B(p_2)
-{1 \over 2}{\int}_l \, H_{\mu}^{\mu} (p_1)
\, B(p_2) \,  + \cr
&+\, 2 i \, {\int}_l \, G_{\mu} (p_1) \, (p_2 - {Q \over 2})_{\mu}
\, B(p_2) \, + \, 3\, {\int}_l B(p_1) \, S(p_2).
\cr
}
\eqno\eq
$$

For all levels $N \ge 1$ one can check that the auxiliary
fields are annihilated (and only them) by the $c_0$ operator.
They play the role of the Lagrange
multipliers. The
rest of the fields are dynamical. Among them are the Stueckelberg fields
which appear as the coefficient functions
corresponding to the 'dynamical' ghost excitations, i.e. the ghost excitations
annihilated by the $b_0$ operator. Such fields are essential for the
construction of the {\it local} gauge invariant theory.
For the first time in our construction they appear on the second level,
namely, the field $S(p)\, $. They are present on all of the higher levels.

As a final comment, note that the physical spectrum of the $2d$ string
theory (the tachyon and the discrete states) can be naturally
reproduced in the
second quantized language ([27]). This boils down to solving
the classical EOM for the free theory $Q \, A = 0$, modulo gauge
invariances $Q \Lambda$.
The
tachyon survives intact, since there is no gauge transformations
associated to that field. All higher order fields are subject to gauge
fixing which kills all but the discrete degrees of freedom (naive
counting of the $2d$ photon degrees of freedom, eg., gives $2 \, -
2=\, 0$). That the states obtained that way are precisely the same as
the ones obtained in the first quantized approach is not surprising,
of course, since we are solving the same set of equations.
What is less obvious, however,
is the question what happens to the discrete states upon including
the interaction. This will be discussed elsewhere. From that end, let
us just note that in the field theoretical
formalism, such a question
is a perfectly well defined one.

\section {Gauge Fixing and the Feynman Rules}

In the Sec.4.1 we were discussing the classical SFT.
Now, the time has come to quantize it. That is to say, we want
to calculate the path integral:

$$
\int \, [d \,a_s] e^{\, -W_{cl}\, (a_s)}   \, ,
$$

\noindent
where $W_{cl}$ is discussed in detail in the previous section.
Such an integral is an ill--defined object, (due to the gauge invariance
$(3.5)$), so
it cannot serve as a starting point for the perturbation theory.
Standard way out in the gauge theories is to fix the gauge. We choose
the Siegel's gauge $b_0 {\ket A} =0$.
After performing the Faddeev - Popov trick once we are left, in general,
with some
gauge condition(s) imposed on the {\it
quantum} fields and with the Jacobian (Faddeev--Popov determinant)
which can be represented as a path integral
over ghosts. In an irreducible theory, such as the Yang Mills, that proves
to be
enough, and from such an effective action one can read off the
propagators and the vertices. In the case of the string theory, which is
an infinitely reducible theory, such a gauge fixed action has additional gauge
invariances. So we have to gauge fix again. By doing so, we
introduce ghosts--of--ghosts. This procedure continue {\it ad
infinitum} [15]. It is convenient, following Thorn, to take the
input (classical) field and all ghosts--of ghosts ... --of ghosts to
be odd. Then, the proper gauge fixed action is:
$$
W_{G.F.} = \, {1 \over 2} \, \int \, (\, A \star Q \, A \, + \, {2 \over
3} \, A \star A \star A \, -2 \, (b_0 \beta) \star A \, )
\eqno\eq
$$
\noindent
where the field $A$ now is the sum of the input field $A_{inp}$
{\it and} all of the ghost fields. It contains fields of the all
possible ghost numbers. The
fact that $\ket A$ is odd means that if $g_s$ is even, the
corresponding coefficient function $a_s$ should be Grassman odd and vice versa.
The field $\beta$ is a Lagrange multiplier (it also contains all
possible ghost numbers) which enforces the gauge condition.

In order to get the Feynman rules, let us integrate over the
$\beta$ fields. We get $b_0 {\ket A} = 0$, or, using
$(3.1)$, $b_0 \ket s = 0$. The kinetic term becomes:
$$   \eqalign
{
&{1 \over 2}\int A \star Q \, A \, = {1 \over 2}\, \int \, A \star Q \,
b_0\,  c_0 \, A \,=
\cr
&= {1 \over 2}\, \int A \, \star (\,  -b_0 \, Q \, c_0 A) + {1 \over 2}\,
\int A \,
\star c_0 \, (L_0 -1) \, A\, =
\cr
&= {1 \over 2}\, \int \, b_0 A \star Q \, c_0 \, A \, + {1 \over 2}\, \int \,
A \star c_0
\, (L_0 -1) \, A \, =
\cr
&= {1 \over 2}\, \int \, A \star \, c_0 \, (L_0 - 1) \, A \, .
}
\eqno\eq
$$
\noindent
One can rewrite $(4.9)$ in terms of the components
fields:
$$  \eqalign
{
&{1 \over 2} \, \int \, A \star c_0 \, (L_0 - 1) \, A \, =
\,{1 \over 2} \sum_{s,l} \, K_{s l}
\, a_{l} \, a_{s} \, ,   \cr
&{\rm where}   \cr
&K_{s l} \, \e \, {}_{21}{\bra V} {\ket s}_1 \, c_0^2 \, (L_0 - 1)
\, {\ket l}_2  \,  .
}
\eqno\eq
$$
\noindent
To add more 'meat' to this rather abstractly looking expression let us
analyse its content on the first couple of levels. One
naive guess would be to just exclude from the input action all of the
auxiliary fields (or, in other words, to put $\Psi$, $B$, $B_{\mu}$,
etc equal to zero). Although this is not the complete answer, as one may
guess, such 'short'
gauge fixed action contains all the information necessary for the tree
amplitude calculations. Nevertheless, it is instructive to see the whole
structure. In fact, $\ket A = {\ket A}_{inp} + {\ket A}_{gh}$, where,
to the second level, ${\ket A}_{gh} =  - i b_{-1} {\ket p} \beta (p) \,
+\, b_{-2} {\ket p} \delta (p) \, -i \am_{-1}
b_{-1} {\ket p} {\beta}_{\mu} \, - i c_{-2} {\ket p} \rho + \am_{-1} c_{-1}
{\ket p} {\gamma}_{\mu} \,$. Here, in line with our conventions,
the coefficient functions
corresponding to the ghost numbers $g=0$ and $g=2$ are Grassmann odd.
One sees that the $g=0$ fields (ghosts) are of the form
$\Lambda \cdot \theta$ where $\theta$ is a Grassmann number and
$\Lambda$ is a gauge parameter field. Ghost number $2$ fields
correspond to the antighosts. The kinetic
part of the action becomes:
$$  \eqalign
{
&{1 \over 2} \int A \star c_0 ( L_0 -1 ) A \, = \, {1 \over 2}\, {\int}_l
\, \Phi \,
(\kin \, ) \, \Phi \, + \, {1 \over 2}\, {\int}_l
\, A_{\mu} \,
(\kin \, +1) \, A_{\mu}  \cr
&- {\int}_l
\, S \,
(\kin \, +2) \, S \,
 + \, {1 \over 4}\, {\int}_l
\, H_{\mu \nu} \,
(\kin \, +2) \, H_{\mu \nu} \,+ \cr
&+ \, {\int}_l
\, G_{\mu} \,
(\kin \, +2) \, G_{\mu} \, - i\, \, {\int}_l
\, \beta \,
(\kin \, +1) \, \gamma \, - \, \cr
&-\,i \, {\int}_l
\, {\beta}_{\mu} \, (\kin \, +2) \, {\gamma}_{\mu} \, -
\,i {\int}_l
\, \rho \,
(\kin \, +2) \, \delta \, \e W_{11} \, + \, W_{20}.
}
\eqno\eq
$$
\noindent
In the last line we denoted the matter and the ghost contributions,
respectively,  as $W_{11}$ and $W_{20}$.

Let us proceed to the interactions. Taking into account signs
convention, one has:

$$
{1 \over 3} \, \int \,  (\, A \star A \star A\, ) = {1 \over 3} \,
\sum_{s,l,m} V_{s l m} \, a_m \, a_l \, a_s \, (-)^{a_l} \, ,
\eqno\eq
$$

\noindent
where the vertex functions are given by:
$$
V_{s l m} \, = \, {}_{321}{\bra V} {\ket s}_1 {\ket l}_2 {\ket m}_3 \,
{}.
\eqno\eq
$$
\noindent
Inserting the coupling constant $g$ in front of the interaction term,
we get the final expression that we have been looking for:

$$
W_{g.f.} \, = {1 \over 2} \sum_{s,l} \, K_{s l}
\, a_{l} \, a_{s} +\, {g  \over 3} \, \sum_{s,l,m} V_{s l m} \, a_m \, a_l \,
 a_s \, (-)^{a_l}  \, .
\eqno\eq
$$

\noindent
It is important to note that $g = n \ne 1$ fields couple to $g=1$ fields
always in pairs with $g = 2-n$. This means that they can enter the
amplitude only in form of the closed ghost loops. Hence, they do
not contribute to the tree
amplitudes. This proves that our naive guess of
the gauge fixed action is indeed correct on the tree level. In the next
section, we implement these results to get
the leading contributions to the 4- and 5-point tachyon scattering
amplitudes.

\chapter {The Calculation of the Open String Amplitudes in d=2}

Now we can proceed to the calculation of the lowest order
contributions to the tachyon 4- and 5- point
on--shell bulk amplitudes.
An advantage of the component perturbation theory is its
straightforwardness. The
calculations of the correlation functions are very similar to those
of the ordinary cubic scalar field
theory. Let us consider 4-point calculations in more
detail. In that case, apart from an overall multiplicative numerical
factor (which we will be often sloppy about), one has, to the leading
order:
$$
g^2 \, \langle \Phi(k_1) \, \cdots \Phi(k_4) V_{s l k} \, a_k a_l a_s \, V_{u
v n} a_n a_v a_u \, (-)^{a_l + a_v} \rangle \, .
\eqno\eq
$$

\noindent
Using the Wick's theorem, chopping the external tachyonic legs and
concentrating on one particular contribution corresponding to the $s$
channel (other contributions are related to this one by the label
permutations) one gets:
$$
A^{(4)}_s \, \propto \, V_{\Phi I \Phi}(k_2, p, k_1) \, D_{I I} (p,
p') \, V_{\Phi I \Phi} (k_4, p', k_3) .
\eqno\eq
$$

\noindent
Here, $D_{I I} (p,p') \,$ is a propagator of an intermediate (off
shell) state $I$ and the vertices $V_{\Phi I \Phi}(k_2, p, k_1)$ are the
couplings of the two on-shell tachyons coupled to an intermediate state $I$.
There are infinitely many intermediate states which contribute to the
amplitude $(5.2)$, so the {\it total}
amplitude is the sum over all of them. Instead of calculating the
whole sum (after all, this can be more efficiently done by the
conformal mapping method [26]), we are interested in the pole
structure itself. The method is {\it designed} precisely to give that
structure. As a check, we have to compare the residues read off
from $(5.2)$ with the ones calculated from the total amplitude [24]. We
confine ourselves to the first three poles. In that case the
propagators are:
$$  \eqalign
{
&D_{\Phi \Phi} (p_1, p_2) \, = \, {1 \over \kin} \, \delta (p_1 + p_2 + Q) \,
, \cr
&D_{A A}(p_1, p_2) \, = \, {\delta^{\mu \nu} \over {\kin + 1}} \, \delta (p_1 +
 p_2 + Q) \, , \cr
&D_{G G}(p_1, p_2) \, = \,-{1 \over 2} \, {\delta^{\mu \nu} \over {\kin + 2}} \
, \delta (p_1 + p_2 + Q) \, , \cr
& D_{S S}(p_1, p_2) \, = \,- {\delta^{\mu \nu} \over {\kin + 2}} \, \delta (p_1
+ p_2 + Q) \, , \cr
&D_{H H}(p_1, p_2) \, = \,{1 \over 4} \, {{\delta^{\mu \kappa} \delta^{
\la \nu} \, + \, \delta^{\mu \la}\delta^{\kappa \nu}} \over {\kin + 2}} \, \
\delta (p_1 + p_2 + Q) \, . \cr
}
\eqno\eq
$$

\noindent
Omitting the common factor of $({4 \over {3
\sqrt3}})^{\kin}\, \delta (k_1 + k_2 + p + Q)$, vertices
$V_{\Phi I \Phi}$ are:

$$   \eqalign
{
&V_{\Phi \Phi \Phi} (k_1, p, k_2) \, = 1 \,  \cr
&V_{\Phi A \Phi} (k_1, p, k_2) \, =  \, N^{12}_{10} \, (k_2^{\mu} \, - k_1^
{\mu})\,  , \cr
&V_{\Phi G \Phi} (k_1, p, k_2) \, = \, 2 \, N^{22}_{20} \, (p^{\mu} +
{1 \over 3}
Q^{\mu}) - \,
 {1 \over 3} Q^{\mu} \,  , \cr
&V_{\Phi H \Phi} (k_1, p, k_2) \, =  \, N^{22}_{11} \, {\delta}^{\mu \nu}
\, +
(N^{12}_{10})^2 \, (k_2 - k_1)^{\mu} (k_2 - k_1)^{\nu} \,  \,
 , \cr
&V_{\Phi S \Phi} (k_1, p, k_2) \, = \, {1 \over 2} + {1 \over 2}
N^{22}_{11} \,    \, . \cr
}
\eqno\eq
$$

\noindent
Plugging $(5.3)$ and $(5.4)$ into $(5.2)$ and integrating out the
$\delta$ functions, one gets:
$$  \eqalign
{
&A^{(4)}_s \, \propto \, g^2 \, ({16 \over 27})^{{1 \over 2} (k_1 +
k_2 + {Q \over 2})^2}  \, ( {1 \over {{1 \over 2} (k_1 +
k_2 + {Q \over 2})^2}} \, + \,{{{4 \over 27}(k_1 - k_2)(k_3 - k_4)}
\over {{1 \over 2} (k_1 +
k_2 + {Q \over 2})^2 \, +1}}\,+\cr
& \cr
&+ \, {{{2 \over 81} (k_1 + k_2 + Q)^2 \, +
{8 \over 729} (k_1 - k_2)(k_3 - k_4)} \over  {{1 \over 2} (k_1 +
k_2 + {Q \over 2})^2 \, +2}}\, + \cr
&+\, {{{124 \over 729} - \, {10 \over 729}
 ( (k_1 -
k_2)^2 \, + \,(k_3 -
k_4)^2 )} \over {{1 \over 2} (k_1 +
k_2 + {Q \over 2})^2 \, +2}}\,+ \cdots  ) \, \delta (\,
\sum_{i=1}^4 k_i\, +\, Q)
.  \cr
}
\eqno\eq
$$

\noindent
Written in the form $(5.5)$ the amplitude seems to have multiparticle poles
(poles in more than one variable). It is not,
however, the case. The reason for this is the $\delta$ function in
$(5.5)$. In fact, upon introducing the 'mass-like' variables $m_i = -{1
\over 2} k_i^2$, and, for definitness, choosing the kinematic region to
be $(+\, +\,  +\,  -)$, $(5.3)$ becomes:
$$  \eqalign
{
&A_{+++-} \, \propto \, g^2 \, ({16 \over 27})^{m_3} ( \, {1 \over
m_3} \, + \, {{{8 \over 27}(m_2 - m_1)} \over {m_3 + 1}} \, + \cr
&+\, {{{120 \over 729} \, + \, {76 \over 729} m_3 \, + \, {32 \over 729}
(m_1 - m_2)^2} \over {m_3 + 2}} \, + \cdots ) \, . \cr
}
\eqno\eq
$$

\noindent
We see that, indeed, four point amplitudes exhibit poles for the
discrete values of the individual momenta only. Calculating
the residues (here we use the kinematic relations $(5.14)$ for $n=3$,
$m=1$) one finally obtains:

$$  \eqalign
{
&A_{+++-} \, \propto \, g^2 ( {1 \over m_3} \,+ \, {1 \over 2} {{(m_2
- m_1)} \over {m_3 + 1}} \, + \cr
&+\, {{1 - {1 \over 2} m_1 m_2} \over {m_3 + 2}} \, + \cdots )\, .
\cr
}
\eqno\eq
$$

\noindent
Five particle amplitudes can be calculated in the similar fashion.
They are proportional to the:
$$  \eqalign
{
&A \, \propto \, g^3 \, V_{\Phi I \Phi} (k_2, p, k_1) D_{I I} (p_1,
p_1') V_{I \Phi J} (p_1', k_3, p_2)  \times \cr
&\times \, D_{JJ}(p_2, p_2') V_{\Phi J \Phi} (k_5, p_2', k_4) \, . \cr
}
$$

\noindent
Thus, to  calculate 5--point amplitudes, one needs to know, also,
the couplings of an on--shell tachyon to $2$ intermediate off--shell particles,
$V_{I \Phi J}$, (we again omit the common factors of
$({4 \over {3 \sqrt3}})^{\sum_{i=1}^2 \, {(p_i + {Q \over
 2})^2 \over 2}} \, \delta \, (\sum_{i=1}^2 \, p_i \,+\, k\,  + Q)$ ):
$$   \eqalign
{
&V_{\Phi \Phi \Phi} (p_1, k, p_2)\, = \, 1 \, , \cr
&V_{\Phi \Phi A} (p_1, k, p_2)\, = \, N^{12}_{10} \, (p_1 \, -k)^{\mu}
\, , \cr
&V_{A \Phi A} (p_1, k, p_2)\, = \,N^{12}_{11} \, {\delta}^{\mu \nu}
\, +
(N^{12}_{10})^2 \, (k - p_2)^{\mu} (p_1 - k)^{\nu} \, . \cr
}
\eqno\eq
$$

\noindent
In the kinematic region $(4,1)$ the amplitude reads:

$$  \eqalign
{
&A_{++++-} \, \propto \, g^3 \, ({16 \over 27})^{(1 - (m_1 + m_2))} \,
({16 \over 27})^{2 m_4} \, \cdot
(\, {1 \over {1 - (m_1 + m_2)}} {1 \over {2 m_4}} \,+\cr
&+ \,
(N^{12}_{10})^2 (\, {{3 (m_3 - m_1 - m_2) + 1 - m_4} \over {(1 -(m_1 +
m_2))(2 m_4 + 1)}} \, + \,  \, {{2 (m_2 - m_1)}
\over {2 m_4 (2 - (m_1 + m_2))}}\, )
\, +\cr
&+ \, 3 (N^{12}_{11} - 4 (N^{12}_{10})^2) \, {{(m_2 - m_1)} \over {(2m_4 +
1)(2 - (m_1 + m_2))}} \, + \, \cdots \,)\, . \cr
}
\eqno\eq
$$

\noindent
Note that the kinematic constraints $(5.14)$ in this case are not
powerful enough to eliminate the two--particle poles, as it was the
case for the 4-point amplitude. It is easy
to see that it is a generic trait, for all $N = n+m \ge 5$. Another interesting
property of the amplitude $(5.9)$ is the presence of the {\it fake}
poles in it. The point is that apart from the poles which we expect,
there are also, on the first sight, poles for the half integer values
of $m_4$. As one can easily check, however, their residues {\it
vanish}. This follows from the kinematics as well as from the
properties of the
Neumann functions. As an example let us state that $N^{12}_{11} -
4 (N^{12}_{10})^2 = 0$ so that the 4th term in $(5.9)$ vanishes even
before taking the residue. Second pole, however, vanishes only after
the residue is taken. One can check that generically, both types of
cancellations appear for the higher intermediate states. After taking
the residues, one gets:

$$
A_{++++-} \, = \, g^3 \, ({1 \over {1-(m_1+m_2)}} {1 \over m_4} \, +
{1 \over 2} {{(m_2 - m_1)} \over {m_4 (2-(m_1+m_2))}}\,+ \cdots ) \, .
\eqno\eq
$$

Let us now briefly summarize the relevant results of Bershadsky and
Kutasov [24]. One denotes a tachyon on-shell vertex operator as
$T^{(\pm)}_k$, $T^{(\pm)}_k = \int _{\partial} \, d \xi e^{i k_{\pm}
\cdot X}$, where:

$$
k^{\mu}_{\pm} \, = \, (k, \, -{Q \over 2} \pm k) .
\eqno\eq
$$

\noindent
The signs '$+$' or '$-$' in $(5.11)$ correspond to the different
chiralities. Correlation functions for the finite boundary and
bulk cosmological constant are not known. The bulk amplitudes, which satisfy
the condition $\sum_{i=1}^N \, k^{\mu}_i + Q^{\mu} \, = \, 0$, can be
calculated explicitly, however. Namely, one finds that in that case:

$$   \eqalign
{
A_{open} (k_1, \dots , k_N) \, = &\propto \int_1^{\infty} \, d
{\xi}_{N-1} \, \int_1^{\xi_{N-1}} \, \cdots \, \int_1^{\xi_4} \, d
\xi_3  \, \cr
&\langle T_{k_1}(0) \, T_{k_2}(1) \, T_{k_1}(\xi_3)\, \cdots T_{k_{N-
1}}(\xi_{N-1}) T_{k_N}(\infty) \rangle   \, . \cr
}
\eqno\eq
$$

\noindent
Upon performing the integrals $(5.12)$ one sees that the result depends on the
kinematics of the tachyons. The only, in
principle, nonvanishing amplitudes are of the type:

$$  \eqalign
{
&A^{(n,m)} \, = \, \langle T^{(+)}_{k_1} \, \cdots \, T^{(+)}_{k_n}
\, T^{(-)}_{k_{n+1}} \, \cdots \,  T^{(-)}_{k_{n+m}} \rangle \, = \cr
&\prod_{i=1}^{n+m} \, {1 \over {\Gamma (1 - m_i)}} \, F_n (k_1, \dots
, k_n) \, F_m (k_{n+1}, \dots , k_{n+m}) \, , \cr
}
\eqno\eq
$$

\noindent
where m's satisfy the kinematic constraints

$$
\sum_{i=1}^n \, m_i \, =
\, 2-m\, , \sum_{i=n+1}^{n+m} \, m_i \, =
\, 2-n\,  ,
\eqno\eq
$$

\noindent
and the form factors $F_n$ are:

$$
F_n \, = \, \prod_{l=1}^{n-1} \, {1 \over {sin \pi \, \sum_{i=1}^l
m_i}} \, .
\eqno\eq
$$

\noindent
To be able to compare $(5.13)$ with $(5.7)$, let us state the explicit form of
$A^{(3,1)}$:

$$ \eqalign
{
&A^{(3,1)} \, \propto \, \prod_{i=1}^4\, {1 \over {\Gamma (1 - m_i)}} \,
{1 \over sin \pi m_1} {1 \over sin \pi (m_1+m_2)}  \, = \cr
&= \, {{\Gamma (m_1) \Gamma (m_3)} \over {\Gamma (1 - m_2)}} \, . \cr
}
\eqno\eq
$$

\noindent
The poles in $m_3$ originate from the $\Gamma (m_3)$. Residue in $m_3
= 0$ is equal to $1$. The residue in $m_3 = -1$ is $(1 - m_1) = {{m_2
- m_1} \over 2}$ (here we have used $m_1 + m_2 = 2$). The next pole has the
residue ${1 \over 2} (1 - m_2) (2 - m_2)$, which can be rewritten as
$(1 - {{m_1 m_2} \over 2})$. One sees that the residues exactly match
the ones in $(5.7)$. It is indeed amusing to see how the messy
coefficients in $(5.5)$
all end up giving simple (and correct) final answer $(5.7)$. Of course such a
good agreement is not a coincidence, since the full answer, for the
four point amplitude, was obtained
in the second quantized framework using the different method [26].
Nevertheless, it is instructive to check the agreement, also,  for the
five point functions.

The five point amplitude $A^{(4,1)}$ is:

$$
A^{(4,1)} \, \propto \, \prod_{i=1}^5 \,{1 \over {\Gamma (1 - m_i)}} \,
{1 \over {sin \pi m_1}} {1 \over {sin \pi (m_1+m_2)}} {1 \over {sin \pi
m_4}}
\eqno\eq
$$

\noindent
Once again, one can see that the five point amplitudes exhibit
multiparticle along with the single particle (discrete) poles.
The inspection shows that $(5.17)$ has the poles in two
independent sets of
variables, eg., $m_4$ and $m_1 + m_2$, for the negative (positive)
integer values, respectively.
Taking the corresponding residues in $(5.17)$ it is easy to confirm
that they give exactly the same pole structure
as $(5.10)$. This, together with the
cancellation of the fake poles in $(5.10)$ clearly indicates that the results
obtained from the perturbation field theory agree with the first quantized
ones.

To conclude, let us outline the relationship between the method used in the
calculations of the correlation functions in this paper
and the one used in [26]. Basically, the conformal mapping method
developed in [16], and used for $2d$ strings in [26]
calculates the amplitude by transforming the field theoretic answer
for the four point on--shell amplitude into an integral which
exactly matches the first quantized expression. In that sense, it is
by construction clear that it reproduces the first quantized
results in a, therefore, somewhat trivial fashion. On the other hand,
the field theory written in components, does not give, as we have seen,
closed expressions for the amplitudes. To get them, one would have to
sum over the perturbation
series. What the method does provide for, however, is the manifest pole
structure of the amplitudes. Origin of poles as coming from the higher
string modes becomes transparent. It is nontrivial check of the
formalism that the residues exactly match ones obtained from [24].

\chapter {Concluding Remarks}

There are several open questions and directions of possible future
investigations. Let us mention here three of them. First, the natural
question arises about the destiny of the discrete states upon
including the interactions (on the classical and the quantum levels).
Second is the important question of constructing the effective tachyonic
action starting from the Witten's SFT for the open strings and,
possibly, connecting it to the collective field theory. Finally, there
is an important problem of how (if at all) cosmological constant can
'arise' from the SFT. These are just some of many interesting
questions one can ask. We hope, and that was the principal aim of this
work, that we have established here the firm ground for these, and
others, future investigations.

\noindent
{\bf Acknowledgments.} I would like to thank A. Jevicki for his
guidance and support throughout the work on this project. I am indebted to B.
Zwiebach and K. Itoh for useful comments on the subject and to I.Ya.
Aref'eva for bringing to my attention the ref. [28].

\vfill
\endpage

\centerline{\it REFERENCES}
\bigskip

\pointbegin
D. J. Gross and A. A. Migdal, {\it Phys. Rev. Lett.} {\bf 64} (1990)
127;
M. R. Douglas and S. Shenker, {\it Nucl. Phys.} {\bf B335} (1990) 635;
E. Br\'ezin and V. Kazakov, {\it Phys. Lett.} {\bf B236} (1990) 144.

\point
D. J. Gross and N. Miljkovi\'c, {\it Phys. Lett.} {\bf B238} (1990) 217;
E. Br\'ezin, V. A. Kazakov and A. B. Zamolodchikov, {\it Nucl.Phys.}
{\bf B338} (1990) 673;
P. Ginsparg and J. Zinn-Justin, {\it Phys. Lett.} {\bf B240} (1990) 333;
D. J. Gross and I. R. Klebanov, {\it Nucl. Phys.} {\bf B344} (1990) 475.

\point
S. R. Das and A. Jevicki, {\it Mod. Phys. Lett.} {\bf A5} (1990) 1639;
A. Jevicki and B. Sakita, {\it Nucl.Phys.} {\bf B165} (1980) 511.

\point
J. Polchinski, {\it Nucl.Phys.} {\bf B362} (1991) 25.

\point
D. Gross and I. Klebanov, {\it Nucl. Phys.} {\bf B352} (1991) 671;
A. M. Sengupta and S. Wadia, {\it Int. J. Mod. Phys.} {\bf A6} (1991)
1961;  G. Moore, {\it Nucl. Phys.} {\bf B368} (1992) 557.

\point
K. Demeterfi, A. Jevicki and J. P. Rodrigues, {\it Nucl.Phys.}
{\bf B362} (1991) 173; {\bf B365} (1991) 499; {\it Mod. Phys. Lett.}
{\bf A35} (1991) 3199;

\point
P. Di Francesco and D. Kutasov, preprint PUPT-1276;
P. Di Francesco and D. Kutasov, {\it Phys. Lett.} {\bf B261} (1991)
385.

\point
G. Moore and R. Plesser, \lq\lq Classical Scattering in 1+1 Dimensional
String Theory", Yale preprint YCTP-P7-92, March 1992.

\point
I. R. Klebanov, \lq\lq Ward Identities in Two-Dimensional String
Theory", PUPT-1302 (1991);
D. Kutasov, E. Martinec and N. Seiberg, PUPT-1293, RU-31-43.

\point
S. Mukhi and C. Vafa, preprint HUTP-93/A002, TIFR/TH/93-01.

\point
J. Avan and A. Jevicki, {\it Phys. Lett.} {\bf B266} (1991) 35;
{\bf B272} (1992) 17.

\point
E. Witten, {\it Nucl.Phys.} {\bf B268} (1986) 253.

\point
 D. Gross and A. Jevicki, {\it Nucl.Phys.} {\bf B283} (1987) 1;
{\it Nucl.Phys.} {\bf B287} (1987) 225.

\point
E. Cremmer, A. Schwimmer and C. Thorn, {\it Phys. Lett.} {\bf B179}
(1986) 57.

\point
C. Thorn, {\it Nucl.Phys.} {\bf B287} (1987) 61.

\point
S. Giddings, {\it Nucl.Phys.} {\bf B278} (1986) 242.

\point
P. Ramond and M. Ruiz - Altaba, Theoretical Physics Preprint Series
UFTP-87-2.

\point
S. Samuel, {\it Nucl.Phys.} {\bf B308} (1988) 285;
{\it Nucl.Phys.} {\bf B308} (1988) 317.

\point
B. Lian and G. Zuckerman, {\it Phys. Lett.} {\bf B254} (1991) 417;
{\it Phys. Lett.} {\bf B266} (1991) 2.

\point
A. M. Polyakov, {\it Mod. Phys. Lett.} {\bf A6} (1991) 635;
Preprint PUPT (Princeton) -1289 (Lectures given at 1991 Jerusalem
Winter School);
D. Kutasov, "Some Properties of (non) Critical Strings",
PUPT-1272, 1991.

\point
E. Witten, {\it Nucl. Phys.} {\bf B373} (1992) 187;
I. Klebanov and A. M. Polyakov, {\it Mod. Phys. Lett} {\bf A6}
(1991) 3273;
N. Sakai and Y. Tanii, {\it Prog. Theor. Phys.} {\bf 86} (1991) 547;
Y. Matsumura, N. Sakai and Y. Tanii, TIT (Tokyo) -HEEP 127, 186 (1992).

\point
E. Witten, B. Zwiebach, preprint IASSNS-HEP-92/4 (1992).

\point
P. Bouwknegt, J. Mc. Carthy and K. Pilch, CERN-TH.6162/91 (1991);
CERN-TH.6279/91 (1991).

\point
M. Bershadsky and D. Kutasov, PUPT-1283, HUPT-91/A047;
PUPT-1315, HUPT-92/A016.

\point
C. Preitschopf and C. Thorn, {\it Nucl.Phys.} {\bf B349} (1991) 132.
A. Sen, {\it Int. J. Mod. Phys.} {\bf A7} (1992).

\point
I. Ya. Aref'eva and A. V. Zubarev, {\it Mod. Phys. Lett.} {\bf A2}
(1992) 677.

\point
N. Sakai and Y. Tanii, preprint TIT/HEP-203, STUPP-92-129.

\point
I. Ya. Aref'eva and A. V. Zubarev, Interaction of $d=2$ $c=1$ Discrete
States from String Field Theory (preprint SMI).

\point
B. Zwiebach, preprint IASSNS-HEP-92/41, MIT-CTP-2102.

\endpage

\end